%% file: main.tex
\documentclass{jfm}
\pdfoutput=1
\usepackage{graphicx}
\usepackage{amsmath}
\usepackage{mathtools}
\usepackage{esint}
\usepackage{xcolor}
\usepackage{tikz}

\usepackage{hyperref}
\hypersetup{
    colorlinks = true,
    urlcolor   = blue,
    citecolor  = blue,
    linkcolor  = blue,
}

\newcommand*\Eval[3]{\left.#1\right\rvert_{#2}^{#3}}
\newcommand{\uvec}[1]{\boldsymbol{\hat{\mathbf{#1}}}}
\newcommand\Ra{\mbox{\textit{R}}}  
\newcommand{\abs}[1]{\left\vert #1 \right\vert}

\newcommand{\quinopt}{{\sc quinopt}}

\definecolor{matlabblue}{RGB}{0,113,188}
\definecolor{matlabred}{RGB}{216,82,24}
\definecolor{mygrey}{rgb}{0.7,0.7,0.7}
\definecolor{matlabgreen}{rgb}{0.4660    0.6740    0.1880} 
\definecolor{matlabyellow}{rgb}{0.93      0.69      0.13} 
\definecolor{colorbar1}{rgb}{1.000000,0.909091,0.000000}
\definecolor{colorbar2}{rgb}{1.000000,0.818182,0.000000}
\definecolor{colorbar3}{rgb}{1.000000,0.727273,0.000000}
\definecolor{colorbar4}{rgb}{1.000000,0.636364,0.000000}
\definecolor{colorbar5}{rgb}{1.000000,0.545455,0.000000}
\definecolor{colorbar6}{rgb}{1.000000,0.454545,0.000000}
\definecolor{colorbar7}{rgb}{1.000000,0.363636,0.000000}
\definecolor{colorbar8}{rgb}{1.000000,0.272727,0.000000}
\definecolor{colorbar9}{rgb}{1.000000,0.181818,0.000000}
\definecolor{colorbar10}{rgb}{1.000000,0.090909,0.000000}
\definecolor{colorbar11}{rgb}{1.000000,0.000000,0.000000}
\definecolor{colorbar12}{rgb}{0.909091,0.000000,0.000000}
\definecolor{colorbar13}{rgb}{0.818182,0.000000,0.000000}
\definecolor{colorbar14}{rgb}{0.727273,0.000000,0.000000}
\definecolor{colorbar15}{rgb}{0.636364,0.000000,0.000000}
\definecolor{colorbar16}{rgb}{0.545455,0.000000,0.000000}
\definecolor{colorbar17}{rgb}{0.454545,0.000000,0.000000}
\definecolor{colorbar18}{rgb}{0.363636,0.000000,0.000000}
\definecolor{colorbar19}{rgb}{0.272727,0.000000,0.000000}
\definecolor{colorbar20}{rgb}{0.181818,0.000000,0.000000}
\definecolor{colorbar21}{rgb}{0.090909,0.000000,0.000000}
\definecolor{grey}{rgb}{0.7,0.7,0.7}

\newcommand\solidrule[1][10pt]{\rule[0.5ex]{#1}{1.5pt}}
\newcommand\dashedrule{\mbox{%
		\solidrule[2pt]\hspace{2pt}\solidrule[2pt]\hspace{2pt}\solidrule[2pt]}}

\newcommand\dottedrule{\mbox{%
		\solidrule[1pt]\hspace{1pt}\solidrule[1pt]\hspace{1pt}\solidrule[1pt]\hspace{1pt}\solidrule[1pt]\hspace{1pt}\solidrule[1pt]\hspace{1pt}}}

\title{Bounds for internally heated convection with fixed boundary heat flux}

\author{Ali Arslan\aff{1}
  \corresp{\email{a.arslan18@imperial.ac.uk}},
  Giovanni Fantuzzi\aff{1},
  John Craske\aff{2}
 \and Andrew Wynn\aff{1}}

\affiliation{\aff{1}Department of Aeronautics, Imperial College London, SW7 2AZ, UK
\aff{2}Department of Civil and Environmental Engineering, Imperial College London, SW7 2AZ, UK }

\begin{document}
\maketitle

\begin{abstract}
We prove a new rigorous bound for the mean convective heat transport $\langle w T \rangle$, where $w$ and $T$ are the nondimensional vertical velocity and temperature, in internally heated convection 
between an insulating lower boundary and an upper boundary with a fixed heat flux. 
The quantity $\langle wT \rangle$ is equal to half the ratio of convective to conductive vertical heat transport, and also to $\frac12$ plus the mean temperature difference between the top and bottom boundaries.
An analytical application of the background method based on the construction of a quadratic auxiliary function yields $\langle w T \rangle \leq \tfrac{1}{2}\big(\tfrac{1}{2}+ \tfrac{1}{\sqrt{3}} \big)  - 1.6552\, \Ra^{-\frac13}$ uniformly in the Prandtl number, where \Ra\ is the nondimensional control parameter measuring the strength of the internal heating. Numerical optimisation of the auxiliary function suggests that the asymptotic value of this bound and the $-1/3$ exponent are optimal within our bounding framework. This new result halves the best existing (uniform in $\Ra$) bound \citep[][\emph{Springer}, Table 1.2]{goluskin2016internally} and its dependence on $\Ra$ is consistent with previous conjectures and heuristic scaling arguments. 
Contrary to physical intuition, however, it does not rule out a mean heat transport larger than $\frac12$ at high $\Ra$, which corresponds to the top boundary being hotter than the bottom one on average.
\end{abstract}


\input{revised_versions/final_version}

\input{main.bbl}

\end{document}

%% file: revised_versions/final_version.tex
\vspace*{-6ex}
\section{Introduction}\label{sec:intro}
\noindent
Convection driven by internally generated heat is a common physical phenomenon and underpins several problems in geophysics, such as mantle convection \citep{schubert2001mantle,MantleConvectioninTerrestrialPlanets}.
One important open problem is to characterize the vertical heat transport as a function of the heating strength, measured by the nondimensional Rayleigh number \Ra. Simulations and experiments \citep{hewitt1980large,ishiwatari1994effects,Lee2007,Goluskin2015} reveal that the heat transport increases with the heating strength and heuristic scaling laws based on physically reasonable, but unproven, assumptions have been put forward~\citep{goluskin2016internally}.  However, corroborating or disproving such heuristic arguments through the derivation of rigorous \Ra-dependent bounds remains a challenge \citep{arslan2021IH1}.

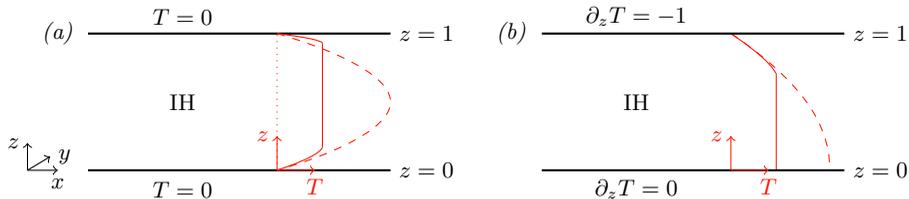
\begin{figure}
    \centering
    \begin{tikzpicture}[every node/.style={scale=0.95}]
    \draw[black,thick] (-6,0) -- (-2,0) node [anchor=west] {$z=0$};
    \draw[black,thick] (-6,1.8) -- (-2,1.8) node [anchor = west] {$z=1$};
    \draw [dashed,colorbar10] plot [smooth, tension = 1] coordinates {(2.5,1.8) (3.5,0.9) (3.8,0)};
    \draw [colorbar10] plot [smooth, tension =1] coordinates {(2.5,1.8) (2.9,1.5) (3.1,1.27)};
    \draw [colorbar10] (3.1,1.27) -- (3.1,0);
    \node at (1.25,-0.25) {$ \partial_z T = 0 $};
    \node at (1.25,2.025) {$ \partial_z T = -1 $};
    \node at (-4.75,0.9) { IH };
    \draw[->,colorbar10] (-3.5,0) -- (-3.5,0.45) node [anchor = east]{$z$};
    \draw[->,colorbar10] (-3.5,0) -- (-3,0) node [anchor=north]{$T$};
    \draw[->] (-6.8,0) -- (-6.8,0.36) node [anchor=east]{$z$};
    \draw[->] (-6.8,0) -- (-6.4,0) node [anchor=north]{$x$};
    \draw[->] (-6.8,0) -- (-6.5,0.18) node [anchor=west]{$y$};
    \draw[black,thick] (0,0) -- (4,0) node [anchor=west] {$z=0$};
    \draw[black,thick] (0,1.8) -- (4,1.8) node [anchor = west] {$z=1$};
    \node at (-4.75,-0.25) {$ T = 0 $};
    \node at (-4.75,2.025) {$  T = 0 $};
    \draw [dashed,colorbar10] plot [smooth, tension = 1] coordinates {(-3.5,1.8) (-2,0.9) (-3.5,0)};
    \draw [colorbar10] plot [smooth,tension=1] coordinates {(-3.5,1.8) (-3.05,1.73) (-2.9,1.67)};
    \draw [colorbar10] (-2.9,1.67) -- (-2.9,0.315);
    \draw [colorbar10] plot [smooth, tension =1] coordinates {(-2.9,0.315) (-3.05,0.18) (-3.5,0)};
    \draw [dotted,colorbar10] (-3.5,0) -- (-3.5,1.8) ;
    \draw[->,colorbar10] (2.5,0) -- (2.5,0.45) node [anchor = east]{$z$};
    \draw[->,colorbar10] (2.5,0) -- (3,0) node [anchor=north]{$T$};
    \node at (1.25,0.9) {\textrm{IH}};
    \node at (-6.4,1.8) {\textit{(a)}};
    \node at (-0.4,1.8) {\textit{(b)}};
    \end{tikzpicture}
    \caption{IH convection with \textit{(a)} isothermal boundaries, studied by \cite{arslan2021IH1}, and \textit{(b)} with fixed-flux boundary conditions, studied in this paper. In both panels, IH represents the uniform unit internal heat generation. Red lines denote the conductive temperature profiles ({\color{colorbar10}\dashedrule}) and indicative mean temperature profiles in the turbulent regime ({\color{colorbar10}\solidrule}).}
    \label{fig:config}
\end{figure}

For internally heated (IH) convection between isothermal boundaries, a major difficulty in bounding the heat transport is
the subtle interplay between the lower
and upper thermal boundary layers \citep{arslan2021IH1}. In
contrast to Rayleigh-B\'{e}nard convection, for which fixed-temperature or fixed-flux boundary conditions are symmetric and
produce unstable thermal boundary layers, internal heating
produces positive buoyancy that acts in the positive vertical direction
and therefore creates asymmetry in relation to the lower and upper boundaries (see Figure \ref{fig:config}$(a)$). In this regard, the lower thermal boundary layer of IH convection \citep[see, for example,][]{goluskin2016penetrative} has a different character to those found
in Rayleigh-B\'{e}nard convection, and is stably
stratified. 

In this study, we remove the
subtleties associated with the lower boundary by specifying a
zero-flux condition, as illustrated in Figure~\ref{fig:config}$(b)$. The hypothesis behind
this choice is that the resulting problem will  be driven primarily by the
properties of the unstably-stratified thermal boundary layer 
near the top boundary and, therefore, will bear a closer
resemblance to Rayleigh-B\'{e}nard convection. To ensure that the energy generated internally leaves the domain and the fluid's temperature does not increase without bound, we also replace the isothermal top boundary with one satisfying a fixed-flux condition.
These boundary conditions idealise models of mantle convection, where radioactive decay provides the internal heating, the core-mantle boundary is approximated by a thermal insulator and a warm crust or atmosphere limit the rate of heat loss to space \citep{trowbridge2016vigorous,MantleConvectioninTerrestrialPlanets,kiefer2009mantle}.

Within this flow configuration, our goal is to bound the dimensionless convective heat flux
$\langle w T \rangle$, where angled brackets denote an average
over volume and infinite time. This quantity is related to the mean temperature difference between the top and bottom boundary: multiplying the equation governing the evolution of temperature (see \S\ref{sec:model}) by the vertical coordinate $z$ and integrating by parts over the volume and infinite time yields
\begin{equation} 
\langle w T \rangle+\overline{T}_{0}-\overline{T}_{1}=\frac{1}{2},
\label{e:J_flux}
\end{equation}
where $\overline{T}_{0}$ and $\overline{T}_{1}$ are the average temperatures of the bottom and top boundaries, respectively, where the average is over the horizontal directions and infinite time. The right-hand side of~\eqref{e:J_flux}, represents the input of potential energy ($1/2$), which balances the reversible work $\langle wT\rangle$, done by the velocity field (equal to the average viscous dissipation) and the unknown rate $\overline{T}_{0}-\overline{T}_{1}$ at which the fluid's potential energy decreases due to conduction. Thus, $\langle wT \rangle=0$ corresponds to the static case of upward conductive transport, $\langle wT \rangle = \tfrac12$ corresponds to purely convective transport between boundaries of equal
mean temperature, and $\langle wT \rangle> \tfrac12$ implies downward conduction on average.

The sign of the conductive term
$\overline{T}_{0}-\overline{T}_{1}$ is a priori unknown, but it can be
shown  that
$|\overline{T}_{0}-\overline{T}_{1}|\leq |\langle T\rangle -
\overline{T}_{1}|^{1/2}$ \citep{goluskin2016internally}. For sufficiently large Rayleigh numbers, this estimate can be combined with the lower bound
$c\Ra^{-1/3}< \langle T \rangle-\overline{T}_{1}$ \citep{lu2004bounds} and the upper bound $\langle T \rangle-\overline{T}_{1} \leq \frac{1}{3}$ \citep{Goluskin2015} to find
$\langle wT \rangle\leq \frac{1}{2}+\frac{1}{\sqrt{3}}$ uniformly in $R$. 
However, assuming that $\overline{T}_{0}$ and $\langle T \rangle$ scale similarly with the Rayleigh number, \cite{goluskin2016internally} conjectured that the mean vertical heat flux should satisfy 
$\langle wT \rangle \leq \tfrac{1}{2} - O(R^{-1/3})$. 

The present work proves that
\begin{equation}\label{e:bound-intro}
    \langle wT \rangle \leq \tfrac12 \left(\tfrac{1}{2}+\tfrac{1}{\sqrt{3}} \right) - c\Ra^{-\frac13},
\end{equation}
with $c \approx 1.6552$. This bound scales with \Ra\ exactly as conjectured and asymptotes to (approximately) $0.5387$, which is slightly larger than $1/2$ but halves the only existing uniform bound. 
To obtain~\eqref{e:bound-intro}, we employ the background method \citep{doering1994variational,doering1996variational,constantin1995variational} interpreted as the search for a quadratic auxiliary function
\citep{chernyshenko2014polynomial,Fantuzzi2016siads,chernyshenko2017relationship,rosa2020optimal}. This interpretation makes it easier to derive a convex variational problem that yields bounds on $\langle w T \rangle$ even though, contrary to traditional applications of the background method, the heat flux in IH convection is not directly related to the thermal dissipation.

The work is structured as follows. Section~\ref{sec:model} presents the governing equations. In \S\ref{sec:fomulating_bound}, we derive the variational problem to bound $\langle wT \rangle$ from above. Analytical and numerical bounds are presented in \S\ref{sec:explicit_ra_bound} and \S\ref{sec:numerical_opt}, respectively. Section~\ref{sec:conclusion} offers concluding remarks.

\vspace*{-3pt}
\section{Model}\label{sec:model} 
\noindent
We consider a uniformly heated layer of fluid bounded between two horizontal plates at a vertical distance $d$. The fluid has kinematic viscosity $\nu$, thermal diffusivity $\kappa$, density $\rho$, specific heat capacity $c_p$, and thermal expansion coefficient $\alpha$. The dimensional heating rate per unit volume $Q$ is constant in time and space. For simplicity, we assume that the layer is periodic in the horizontal ($x$ and $y$) directions with periods $d L_x$ and $d L_y$. While these values affect the mean vertical heat flux, the analytical bounds derived below do not depend on $L_x$ or $L_y$ and, therefore, apply to domains of all sizes (including the limiting case of a horizontally infinite fluid layer).

To make the problem nondimensional, we use $d$ as the characteristic length scale, $d^{2}/\kappa$ as the time scale, and $d^2Q/\kappa \rho c_p$ as the temperature scale.   The motion of the fluid in the nondimensional domain $\Omega = [0,L_x] \times [0,L_y] \times [0,1]$ is then governed by the Boussinessq equations
\begin{subequations}\label{e:governing-equations}
    \begin{align}
    \bnabla \cdot \boldsymbol{u} &= 0\, , \label{continuit} \\
    \partial_t \boldsymbol{u}+ \boldsymbol{u}\cdot \bnabla \boldsymbol{u} + \bnabla p &= \Pran ( \bnabla^{2}\boldsymbol{u} + \Ra T \uvec{z})\, , \label{nondim_momentum} \\
    \partial_t T + \boldsymbol{u}\cdot \bnabla T  &= \bnabla^{2}T + 1,
    \label{nondim_energy}
\end{align}
\end{subequations}
where $\boldsymbol{u}$ is the fluid velocity, $p$ is the pressure, and the unit forcing in~\eqref{nondim_energy} represents the nondimensional internal heating rate. The no-slip, fixed-flux boundary conditions are expressed by
\begin{subequations}\label{e:boundary-conditions-all}
    \begin{gather}
	\Eval{\boldsymbol{u}}{z=0}{} = \Eval{\boldsymbol{u}}{z=1}{} =0, \label{bc_velocity}\\
	\Eval{\partial_z T}{z=0}{} = 0, \quad\Eval{\partial_z T}{z=1}{} = -1.  \label{bc_T_IH2} 
    \end{gather}
\end{subequations}
The dimensionless quantities
\begin{equation}
    \Pran = \frac{\nu}{\kappa} \qquad\text{and}\qquad
    \Ra = \frac{g \alpha Q d^{5}}{\rho c_p \nu \kappa^{2}},
\end{equation}
where $g$ is the acceleration of gravity, are the only two nondimensional control parameters. The former is the usual Prandtl number, which measures the ratio of momentum and heat diffusivity. The latter, instead, measures the destabilising effect of internal heating compared with the stabilising effects of diffusion, and may therefore be interpreted as a Rayleigh number.

Since the volume-averaged temperature  $\fint T(\boldsymbol{x},t) {\rm d} \boldsymbol{x}$ is conserved in time, we assume it to be zero without loss of generality. With this extra condition, the governing equations admit the solution $\boldsymbol{u}=0$, $p = \text{constant}$ and $T = -\tfrac{z^2}{2} + \frac16$ at all $\Ra$, which represents a purely conductive state. This state is globally asymptotically stable irrespective of the horizontal periods $L_x$ and $L_y$ if $\Ra < 1429.86$, and it is linearly unstable when $\Ra > 1440$ 
for sufficiently large horizontal periods \citep{Goluskin2015}. Convection ensues above this Rayleigh number for at least one choice of the horizontal periods, and cannot currently be ruled out above the known global stability threshold. We are therefore interested in bounds on $\langle wT \rangle$ that hold for arbitrary $\Ra \geq 1429.86$.

\section{Bounding framework}\label{sec:fomulating_bound}
\noindent
To derive an upper bound on $\langle wT \rangle$, it is convenient to lift the inhomogeneous boundary condition on the temperature by introducing the temperature perturbation
\begin{equation}
\label{e:T_theta_relation}
\theta(\boldsymbol{x},t) = T(\boldsymbol{x},t) + \frac{z^2}{2} -\frac16.
\end{equation}
The heat equation~\eqref{nondim_energy} and boundary conditions~\eqref{bc_T_IH2} show that $\theta$ satisfies 
\begin{subequations}
	\begin{gather}
	\label{eq:nondim_energy_theta}
	\partial_t \theta + \boldsymbol{u}\cdot \bnabla \theta  =  \bnabla^{2}\theta + z w,\\
	\label{boundary_conditions_theta}
	\Eval{\partial_z\theta}{z=0}{} = 0, \quad\Eval{\partial_z\theta}{z=1}{} = 0. 
	\end{gather}
\end{subequations}

To rewrite the heat flux in terms of $\theta$, observe that, by virtue of incompressibility and of the boundary conditions in~\eqref{bc_velocity}, 
the horizontal-and-time average $\overline{w}(z)$ of the vertical velocity $w$ vanishesfor all $z$.
Then,
\begin{equation}\label{e:k-bound-condition}
   \langle w f(z) \rangle= \int_0^1 \overline{w}(z) f(z) \, {\rm d}z = 0,
\end{equation}
for any $z$-dependent function $f(z)$ and, in particular, we conclude that
\begin{equation}\label{e:wT-identity}
\langle wT \rangle = \langle w\theta \rangle.
\end{equation}  

A rigorous upper bound on $\langle w\theta \rangle$ can be derived using the auxiliary functional method~\citep{chernyshenko2014polynomial} with the quadratic auxiliary functional
\begin{equation}
\label{eq:aux_functional}
\mathcal{V}\{\boldsymbol{u}, \theta \} = 
\fint_{\Omega} \frac{a}{2\Pran \Ra} |\boldsymbol{u}|^{2} 
+ \frac{b}{2} |\theta|^2 - \phi(\boldsymbol{x}) \theta - \boldsymbol{\psi}(\boldsymbol{x}) \cdot \boldsymbol{u}
\; \textrm{d}\boldsymbol{x},
\end{equation}
where the nonnegative scalars $a$ and $b$, the function $\phi$, and the vector field $\boldsymbol{\psi}$ are to be optimised in order to obtain the best possible bound. \cite{chernyshenko2017relationship} showed that this is equivalent to employing the background method: 
the profile $\frac{\phi}{b}$ is the background temperature (defined with respect to $\theta$), 
the vector field $\frac{\text{\it Pr} R}{a} \boldsymbol{\psi}$ is the background velocity,
and $a$ and $b$ are the so-called balance parameters. Contrary to the classical background method, there is no need to impose boundary or incompressibility conditions on $\boldsymbol{\psi}$ or $\phi$ when defining $\mathcal{V}$. To simplify the analysis below, however, we take $\boldsymbol{\psi}$ to be incompressible, horizontally periodic, and vanishing at the top and bottom plates. The optimality of these choices can be proved rigorously, but the details are beyond the scope of this work.

Arguments identical to those given by \citet[Appendix A]{goluskin2019ks} show that the auxiliary function~\eqref{eq:aux_functional} may be taken to be invariant under arbitrary horizontal translations and under the ``horizontal flow reversal'' variable transformation
\begin{equation}\label{e:horiz-flow-reversal}
    \begin{pmatrix}\boldsymbol{u}(\boldsymbol{x},t)\\\theta(\boldsymbol{x},t) \\ p(\boldsymbol{x},t)\end{pmatrix}
    \mapsto
    \begin{pmatrix}\mathsf{G} \boldsymbol{u}(\mathsf{G}\boldsymbol{x},t)\\\theta(\mathsf{G}\boldsymbol{x},t) \\ p(\mathsf{G}\boldsymbol{x},t)\end{pmatrix}, \qquad
    \mathsf{G}= \begin{pmatrix}-1 & 0 & 0\\ 0 & -1 & 0\\ 0 & 0 & 1\end{pmatrix}.
\end{equation}
This is because the governing equations~\eqref{continuit}, \eqref{nondim_momentum} and \eqref{eq:nondim_energy_theta} are invariant under the same set of transformations. Invariance under horizontal translation requires $\phi(\boldsymbol{x})=\phi(z)$ and $\boldsymbol{\psi}(\boldsymbol{x})=\boldsymbol{\psi}(z)$. In particular, the incompressibility and no-slip conditions on $\boldsymbol{\psi}$ imply that we must have $\boldsymbol{\psi}(z) = (\psi_1(z), \psi_2(z), 0)$. Invariance under~\eqref{e:horiz-flow-reversal} then requires $\psi_1(z)=\psi_2(z)=0$, so $\boldsymbol{\psi}=0$.

Making these restrictions from now on, it can be shown that $\mathcal{V}\{\boldsymbol{u}(t), \theta(t) \}$ remains uniformly bounded in time along solutions of~\eqref{nondim_momentum} and~\eqref{eq:nondim_energy_theta} for any given initial velocity and temperature. We can therefore use the fundamental theorem of calculus and the governing equations to write
\begin{align}
\langle w \theta \rangle 
&= \limsup_{\tau \rightarrow \infty}\frac{1}{\tau} \int_{0}^{\tau} \fint_{\Omega} w\theta +  \frac{{\rm d}}{ {\rm d}t}\mathcal{V}\{\boldsymbol{u}(t),\theta(t)\}  \; {\rm d}\boldsymbol{x} \textrm{d}t
\nonumber \\
&= \limsup_{\tau \rightarrow \infty}\frac{1}{\tau} \int_{0}^{\tau} \fint_{\Omega} w\theta 
+  \frac{a}{\Pran\Ra}\boldsymbol{u} \cdot \partial_t \boldsymbol{u}
+ b \theta \partial_t \theta  - \phi(z)\partial_t \theta   \; {\rm d}\boldsymbol{x} \textrm{d}t
\nonumber \\
&= U -\liminf_{\tau \rightarrow \infty}\frac{1}{\tau} \int_{0}^{\tau} \mathcal{S}\{\boldsymbol{u}(t),\theta(t) \} \textrm{d}t
\label{e:af-method-steps}
\end{align}
for any constant $U$, where
\begin{equation}
\label{eq:S_functional_constraint}
\mathcal{S}\{\boldsymbol{u},\theta\} = \fint_{\Omega} \frac{a}{R}|\bnabla\boldsymbol{u}|^{2} + b |\bnabla \theta|^2 - (a+1 +bz - \phi' )w\theta  -\phi' \partial_z\theta + U\, \textrm{d}\boldsymbol{x}
\end{equation}
and primes denote derivatives with respect to $z$. The last equality in~\eqref{e:af-method-steps} is obtained after a few integrations by parts that exploit the boundary conditions, incompressibility, and identity~\eqref{e:k-bound-condition}.
The bound $\langle w T \rangle = \langle w\theta \rangle \leq U$ follows from~\eqref{e:wT-identity} and~\eqref{e:af-method-steps} if $\mathcal{S}\{\boldsymbol{u},\theta\}$ is nonnegative for all time-independent velocities $\boldsymbol{u}$ and temperature perturbations $\theta$ that satisfy incompressibility and the boundary conditions in~\eqref{bc_velocity} and~\eqref{boundary_conditions_theta}. Our goal, therefore, is to choose $a$, $b$ and $\phi(z)$ such that this condition holds for the smallest possible~$U$.

To simplify this task, we invoke the horizontal periodicity and expand the velocity and temperature fields using Fourier series,
\begin{equation}\label{e:Fourier}
    \begin{bmatrix}
    \theta(x,y,z)\\\boldsymbol{u}(x,y,z)
    \end{bmatrix}
    = \sum_{\boldsymbol{k}} 
    \begin{bmatrix}
    \hat{\theta}_{\boldsymbol{k}}(z)\\ \hat{\boldsymbol{u}}_{\boldsymbol{k}}(z)
    \end{bmatrix}
    \textrm{e}^{i(k_x x + k_y y)}\, .
\end{equation}
The sum is over wavevectors $\boldsymbol{k}=(k_x,k_y)$ of magnitude $k = \sqrt{k_{\smash{x}}^2 + k_{\smash{y}}^2}$ that are compatible with the horizontal periods $L_x$ and $L_y$. The (complex-valued) Fourier amplitudes $\hat{\theta}_{\boldsymbol{k}}$ and $\hat{\boldsymbol{u}}_{\boldsymbol{k}}=(\hat{u}_{\boldsymbol{k}},\hat{v}_{\boldsymbol{k}},\hat{w}_{\boldsymbol{k}})$ satisfy
\begin{subequations}\label{e:Fourier-bc}
    \begin{gather}
        \hat{u}_{\boldsymbol{k}}(0) = \hat{u}_{\boldsymbol{k}}(1)=\hat{v}_{\boldsymbol{k}}(0) = \hat{v}_{\boldsymbol{k}}(1)=0,\\
        \label{e:Fourier-bc-wk}
        \hat{w}_{\boldsymbol{k}}(0) = \hat{w}_{\boldsymbol{k}}'(0) = \hat{w}_{\boldsymbol{k}}(1) = \hat{w}_{\boldsymbol{k}}'(1)= 0,\\
        \hat{\theta}_{\boldsymbol{k}}'(0)=\hat{\theta}_{\boldsymbol{k}}'(1) = 0.
        \label{e:Fourier-bc-Tk}
\end{gather}
\end{subequations}

After substituting \eqref{e:Fourier} into \eqref{eq:S_functional_constraint}, the Fourier-transformed incompressibility condition 
$ik_x \hat{u}_{\boldsymbol{k}} + ik_y \hat{v}_{\boldsymbol{k}} + \hat{w}_{\boldsymbol{k}}' = 0$ can be combined with Young's inequality 
to estimate
\begin{equation}\label{e:S-estimate}
    \mathcal{S}\{\boldsymbol{u},\theta\} \geq \mathcal{S}_{0}\{\hat{\theta}_0\} + \sum_{\boldsymbol{k}} \mathcal{S}_{\boldsymbol{k}} \{\hat{w}_{\boldsymbol{k}},\hat{\theta}_{\boldsymbol{k}}\},
\end{equation}
where
\begin{subequations}
	\begin{equation}
	\label{eq:S0}
	\mathcal{S}_0\{\hat{\theta}_0\} 
	:= U + \int^{1}_{0} b|\hat{\theta}'_0(z)|^{2} - \phi'\hat{\theta}'_0(z)\, \textrm{d}z\, 
	\end{equation}
	and
	\begin{multline}
	\label{eq:Sk}
	\mathcal{S}_{\boldsymbol{k}}\{\hat{w}_{\boldsymbol{k}},\hat{\theta}_{\boldsymbol{k}}\} 
	:= \int^{1}_{0} \frac{a}{R}\left(\frac{|\hat{w}''_{\boldsymbol{k}}(z)|^2}{k^2} + 2|\hat{w}'_{\boldsymbol{k}}(z)|^2 + k^{2}|\hat{w}_{\boldsymbol{k}}(z)|^2 \right)  +b|\hat{\theta}'_{\boldsymbol{k}}(z)|^{2}
	\\
	 + bk^{2}|\hat{\theta}_{\boldsymbol{k}}(z)|^2 
	- [a+1+bz - \phi'(z)]\hat{w}_{\boldsymbol{k}}(z)^*\hat{\theta}_{\boldsymbol{k}}(z) \, \textrm{d}z\, .
	\end{multline}
\end{subequations}
Standard arguments~\citep[see, e.g.,][]{arslan2021IH1} show that the right-hand side of~\eqref{e:S-estimate} is nonnegative if and only if each summand is nonnegative, and that to check these conditions one can assume that $\hat{w}_{\boldsymbol{k}}$ and $\hat{\theta}_{\boldsymbol{k}}$ are real functions. Thus, the best bound on $\langle w T \rangle$ is found upon solving the optimisation problem
\begin{equation}\label{e:optimization-Fourier}
	\begin{aligned}
	\inf_{U,\phi'(z),a,b} \big\{U:\quad
	&\mathcal{S}_0\{\hat{\theta}_0\} \geq 0 \quad\forall \,\hat{\theta}_0 \text{ s.t. } \eqref{e:Fourier-bc-Tk},\\[-1ex]
	&\mathcal{S}_{\boldsymbol{k}}\{\hat{w}_{\boldsymbol{k}},\hat{\theta}_{\boldsymbol{k}}\} \geq 0 \quad\forall \, \hat{w}_{\boldsymbol{k}},\hat{\theta}_{\boldsymbol{k}} \text{ s.t. }\text{(\ref{e:Fourier-bc}\textit{a,b})}, \; \forall \boldsymbol{k}\neq 0
	\big\}.
	\end{aligned}
\end{equation}
We refer to the condition $\mathcal{S}_{\boldsymbol{k}}\{\hat{w}_{\boldsymbol{k}},\hat{\theta}_{\boldsymbol{k}}\} \geq 0$ as the spectral constraint and consider $\phi'$, rather than $\phi$, as the optimization variable because only the former appears in the problem.

\section{Analytical bound}
\label{sec:explicit_ra_bound}
\noindent
To derive an analytical bound on $\langle wT \rangle$, we begin by observing that
\begin{equation}\label{e:cost-bound}
    \mathcal{S}_0\{\hat{\theta}_0\} = \int^{1}_{0} b\left(\hat{\theta}'_{0}(z) - \frac{\phi'(z)}{2b} \right)^2 - \frac{\phi'(z)^{2}}{4b} + U\, \textrm{d}z \geq  U - \int^{1}_{0}\frac{\phi'(z)^{2}}{4b}~\textrm{d}z,
\end{equation}
so the constraint on $\mathcal{S}_0$ in \eqref{e:optimization-Fourier} is satisfied if we choose
\begin{equation}
    \label{e:U-bound}
    U = \int^{1}_{0}\frac{\phi'(z)^{2}}{4b}\, \textrm{d}z.
\end{equation}
This choice is also optimal because the lower bound in~\eqref{e:cost-bound} is sharp. To see this, let $\hat{\theta}_0$ be such that $\hat{\theta}_0'(z) = \frac{1}{2b}\phi'(z)$ except for boundary layers of width $\epsilon$ near $z=0$ and~$1$, where $\hat{\theta}_0'(z)=0$ to satisfy~\eqref{e:Fourier-bc-Tk}. Then, let $\epsilon \to 0$ and apply Lebesgue's dominated convergence theorem to conclude that $\mathcal{S}\{\theta_0\}$ converges to the right-hand side of~\eqref{e:cost-bound}.

Next, we seek constants $a$ and $b$ and a function $\phi(z)$ that minimize the right-hand side of~\eqref{e:U-bound} whilst satisfying the spectral constraint in~\eqref{e:optimization-Fourier}. The simplest way to ensure this is to set $\phi'(z)=a+1+bz$, because then the only sign-indefinite term in $\mathcal{S}_{\boldsymbol{k}}$ vanishes. This choice yields
\begin{equation}
\langle wT \rangle \leq U = \frac{1}{12} \left[ b + 3(a+1) + \frac{(a+1)^2}{b} \right]\, ,
\end{equation}
which attains the minimum value of $\tfrac{1}{2}\big(\tfrac{1}{2} + \tfrac{1}{\sqrt{3}}\big)$ when $b=\sqrt{3}(a+1)$ and $a=0$.

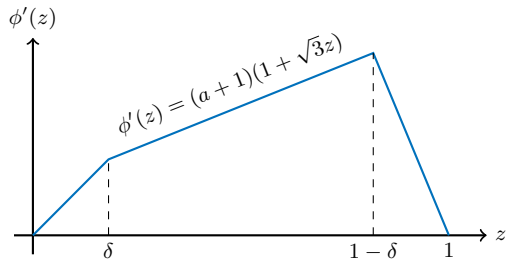
\begin{figure}
    \centering
    \begin{tikzpicture}[every node/.style={scale=0.9}]
    \draw[->,black,thick] (-3.25,0) -- (3,0) node [anchor=west] {$z$};
    \draw[->,black,thick] (-3,-0.25) -- (-3,2.6) node [anchor=south] {$\phi'(z)$};
    \draw[matlabblue,thick] (-3,0) -- (-2,1) -- (1.5,2.4) -- (2.5,0);
    \node[anchor=north] at (2.5,0) {$1$};
    \node[rotate=21.5] at (-0.4,2) {$\phi'(z)=(a+1)(1+\sqrt{3}z)$};
    \draw[dashed] (-2,0) node[anchor=north] {$\delta$} -- (-2,1);
    \draw[dashed] (1.5,0) node[anchor=north] {$1-\delta$} -- (1.5,2.4);
    \end{tikzpicture}
    \caption{Sketch of the piecewise-linear $\phi'(z)$ in~\eqref{eq:phi_g}.}
    \label{fig:phi_g}
\end{figure}

While this simple construction already halves the uniform bound proved by~\cite{goluskin2016internally}, an even better result that depends explicitly on the Rayleigh number can be obtained by letting $\phi'$ develop boundary layers of width $\delta$ near $z=0$ and~$1$. Specifically, we still fix
\begin{equation}\label{e:b-optimal}
b = \sqrt{3}(a+1),
\end{equation}
but this time take
\begin{equation}
\label{eq:phi_g}
\phi'(z) = (a+1)\xi(z), \qquad
\xi(z)= 
\begin{dcases}
\left(\tfrac{1}{\delta}+\sqrt{3}\right)z, &  0\leq z \leq \delta, \\
1 + \sqrt{3}\,z ,  &  \delta \leq z \leq 1-\delta, \\
\left(\tfrac{1 + \sqrt{3}}{\delta}- \sqrt{3}\right)(1-z), & 1-\delta \leq z \leq 1.
\end{dcases}
\end{equation}
This profile, illustrated in Figure~\ref{fig:phi_g}, yields the upper bound
\begin{subequations}
    \label{e:U_explicit_exact_all}
    \begin{align}
        \label{e:U_explicit_exact}
        \langle wT \rangle \leq  
        U &= \frac12  \left( \frac12 + \frac{1}{\sqrt{3}} - \frac{6+5\sqrt{3}}{9}\,\delta + \frac{\sqrt{3}}{6}\,\delta^{2} \right) (a+1)\\
        &\leq \frac12 \left( \frac12 + \frac{1}{\sqrt{3}}  -  A\,\delta  \right) (a+1),
        \label{e:U_explicit}
    \end{align}
\end{subequations}
where the last inequality holds for any constant $A$ satisfying $A \leq  \tfrac{1}{9}(6 + 5\sqrt{3}) - \tfrac{\sqrt{3}}{6}\delta$. Anticipating that the height of the boundary layers in $\phi'$ will have to be small, we arbitrarily assume that $\delta \leq 1/3$ (this will be checked \textit{a posteriori}) and therefore set $A = (4+3\sqrt{3})/6$ irrespective of $\delta$. These conservative choices considerably simplify the algebra in what follows.

Note that although~\eqref{e:U-bound} suggests setting $\xi(z)=0$ throughout the boundary layers, a linear variation makes the spectral constraint easier to satisfy and results in a smaller bound on $\langle wT \rangle$.
The values of $a$ and $\delta$ must be chosen as a function of $\Ra$ to minimize~\eqref{e:U_explicit} whilst ensuring that the indefinite term in $\mathcal{S}_{\boldsymbol{k}}$,
\begin{equation}
     \mathcal{I} := (a+1)\int_{[0,\delta]\cup[1-\delta,1]}[1 + \sqrt{3}z-\xi(z)] w_{\boldsymbol{k}}(z) \theta_{\boldsymbol{k}}(z) \,\textrm{d}z,
\end{equation}
can be controlled. For the boundary layer at $z=0$, we can use the boundary conditions on $w_{\boldsymbol{k}}$ and $w_{\boldsymbol{k}}'$ in~\eqref{e:Fourier-bc-wk} and the Cauchy--Schwarz inequality to estimate
\begin{equation}
    \abs{\hat{w}_{\boldsymbol{k}}(z)} 
    = \abs{\int_0^z \!\! \int_0^\zeta \! \hat{w}_{\boldsymbol{k}}''(\eta) {\rm d}\eta \,{\rm d}\zeta }
    \leq \int_0^z \!\!\int_0^\zeta \! \abs{ \hat{w}_{\boldsymbol{k}}''(\eta)}  {\rm d}\eta \,{\rm d}\zeta
    \leq \int_0^z \!\!\!\sqrt{\zeta} {\rm d}\zeta \, \|\hat{w}_{\boldsymbol{k}}''\|_2
    = \tfrac{2}{3}z^{\frac32}\lVert w_{\boldsymbol{k}}'' \rVert_2.
\end{equation}
Using this estimate, the definition of $\xi$ from~\eqref{eq:phi_g}, and the Cauchy-Schwarz inequality once again we obtain
\begin{equation}
    \abs{\int_0^\delta [1 + \sqrt{3}z-\xi(z)] w_{\boldsymbol{k}}(z) \theta_{\boldsymbol{k}}(z)\,\textrm{d}z}
    \leq \frac{\delta^{2}}{3\sqrt{15}} \lVert w_{\boldsymbol{k}}'' \rVert_{2} \lVert \theta_{\boldsymbol{k}} \rVert_{2}.
\end{equation}
Similar arguments near $z=1$ yield
\begin{equation}
    \abs{\int^{1}_{1-\delta}[1 + \sqrt{3}z-\xi(z)] w_{\boldsymbol{k}}(z) \theta_{\boldsymbol{k}}(z)\,\textrm{d}z}
     \leq  \frac{1+\sqrt{3}}{3\sqrt{15}} \, \delta^{2} \lVert w_{\boldsymbol{k}}'' \rVert_{2} \lVert \theta_{\boldsymbol{k}} \rVert_{2}\,.
\end{equation}
Using these inequalities we can now estimate
\begin{align}
    \mathcal{S}_{\boldsymbol{k}}\{ \hat{w}_{\boldsymbol{k}}, \hat{\theta}_{\boldsymbol{k}}\}
    &\geq \frac{a}{Rk^{2}}\lVert w_{\boldsymbol{k}}'' \rVert_{2}^{2} + \sqrt{3}(a+1)k^{2} \lVert \theta_{\boldsymbol{k}} \rVert_{2}^{2} - \abs{\mathcal{I}}
    \nonumber \\
    &\geq \frac{a}{Rk^{2}}\lVert w_{\boldsymbol{k}}'' \rVert_{2}^{2} + \sqrt{3}(a+1)k^{2} \lVert \theta_{\boldsymbol{k}} \rVert_{2}^{2}
    - \tfrac{2+\sqrt{3}}{3\sqrt{15}}(a+1)\delta^{2}\lVert w''_{\boldsymbol{k}} \rVert_{2} \lVert \theta_{\boldsymbol{k}} \rVert_{2}.
\end{align}
The last expression is a homogeneous quadratic form in $\lVert w_{\boldsymbol{k}}'' \rVert_{2}$ and $\lVert \theta_{\boldsymbol{k}} \rVert_{2}$, and is nonnegative if its discriminant is nonpositive. To ensure that the spectral constraints hold with the largest possible $\delta$, so the bound~\eqref{e:U_explicit} is minimised, we therefore set
\begin{equation}
    \label{e:delta_condition}
    \delta = \alpha\left( \frac{a}{(a+1)R} \right)^\frac14\,,
\end{equation}
with $\alpha = [540(7\sqrt{3} -12)]^{\frac14}$.
Substituting this into \eqref{e:U_explicit_exact} yields an upper bound on $\langle wT \rangle$ that depends only on $a$, and in principle this parameter can be optimized numerically for each value of \Ra. To obtain a fully analytical bound, however, we substitute~\eqref{e:delta_condition} into the weaker bound~\eqref{e:U_explicit} and use the fact that $a>0$ to arrive at
\begin{align}
    \langle wT \rangle
    &\leq \frac12 \left( \frac12 + \frac{1}{\sqrt{3}} \right)(a+1) - \frac{1}{2} A\, \alpha\, a^{\frac{1}{4}} (a+1)^{\frac34} R^{-\frac{1}{4}}
    \nonumber \\
    &\leq \frac12 \left( \frac12 + \frac{1}{\sqrt{3}} \right)(a+1) - \frac{1}{2} A\, \alpha\, a^{\frac{1}{4}}R^{-\frac{1}{4}}.
\end{align}
This bound can be optimised analytically over $a$ by solving the equation $\partial U/\partial a =0$, which gives
\begin{equation}\label{e:a-optimal}
    a = a_0 R^{-\frac13}\,
\end{equation}
with $a_0 = \left[\tfrac{1}{2}A\alpha(2\sqrt{3}-3)\right]^{\frac43}$. This translates into the upper bound
\begin{equation}
    \label{e:explicit_analytic_bound}
    \langle wT \rangle \leq 
    \frac{1}{2} \left( \frac{1}{2} + \frac{1}{\sqrt{3}} \right) \left( 1 
    - 3 a_0 \, R^{-\frac13} \right) \,.
\end{equation}
Substituting~\eqref{e:a-optimal} into~\eqref{e:delta_condition} shows that $\delta = O(R^{-\frac13})$, which agrees with the scaling arguments proposed for Rayleigh-B\'enard convection \citep{spiegel1963generalization}. Moreover, the constraint $\delta \leq \frac13$ imposed at the beginning is satisfied for all $\Ra \geq 591.51$, which is below the energy stability limit (cf. \S\ref{sec:model}). As required, therefore, the upper bound~\eqref{e:explicit_analytic_bound} applies to all values of $\Ra$ for which convection cannot be ruled out.

\section{Numerically optimised bounds}
\label{sec:numerical_opt}
\noindent
To assess how far the analytical bound~\eqref{e:explicit_analytic_bound} is from being optimal, we numerically approximated the best upper bounds on $\langle wT \rangle$ implied by problem~\eqref{e:optimization-Fourier} using the MATLAB\ toolbox \quinopt\ \citep{fantuzzi2017optimization}. This toolbox employs truncated Legendre series expansions for the tunable function $\phi$ and for the unknown fields $\hat{\theta}_0$, $\hat{\theta}_{\boldsymbol{k}}$ and $\hat{w}_{\boldsymbol{k}}$ in order to discretise the convex variational problem~\eqref{e:optimization-Fourier} into a numerically tractable semidefinite program (SDP) \citep[for more details on this approach, see][]{fantuzzi2015construction,Fantuzzi2016PRE,fantuzzi2018bounds}. Numerically optimal solutions to~\eqref{e:optimization-Fourier} were obtained for $10^3 \leq \Ra \leq 10^9$ in a two-dimensional domain with horizontal period $L_x  = 2$. The number of terms in the Legendre series expansion used by \quinopt\ was increased until the optimal upper bound changed by less than $1\%$, and an iterative procedure \citep[see, e.g.,][]{Fantuzzi2016PRE} was employed to check the spectral constraints $\mathcal{S}_{\boldsymbol{k}}\geq 0$ up to the cut-off wavenumber
\begin{equation}
    \label{e:k_cutoff}
    k_c:=\left(\frac{R}{4ab}\right)^{\frac14}\lVert a + 1+ bz - \phi' \rVert^{\frac12}_{\infty}\, ,
\end{equation}
where $\lVert \cdot \rVert_{\infty}$ is the $L^\infty$ norm. This value was derived using the method described in \citet[][Appendix B]{arslan2021IH1}, which ensures that $\mathcal{S}_{\boldsymbol{k}}\geq 0$ is nonnegative for all $k > k_c$ given any fixed choices of $\Ra$, $a$, $b$ and $\phi'$.
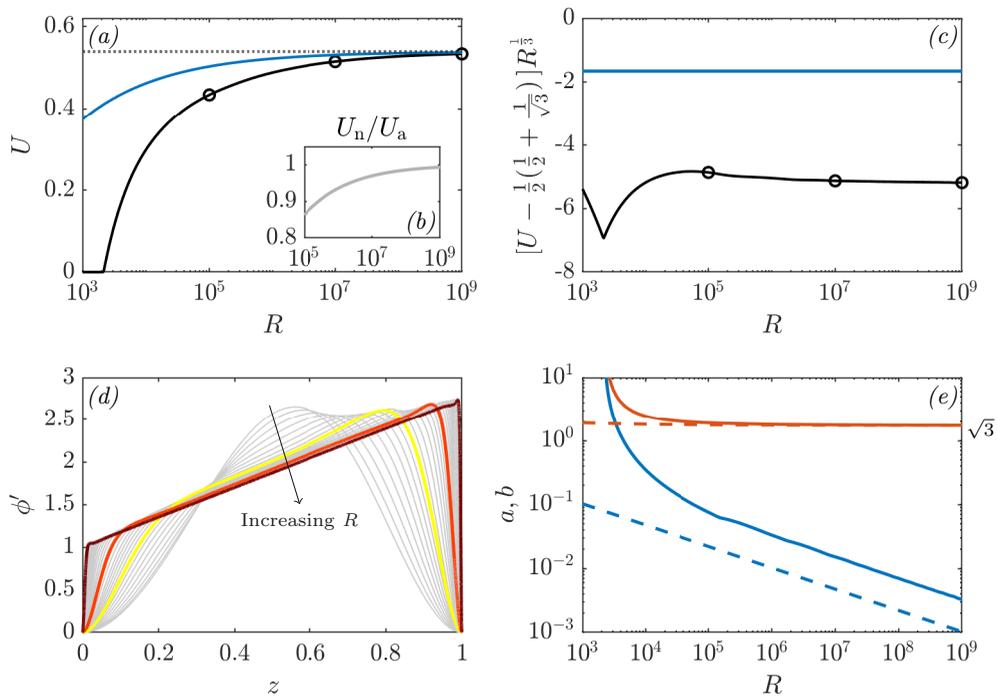
\begin{figure}
    \centering    
    \hspace*{-1cm}
    \includegraphics[scale=1]{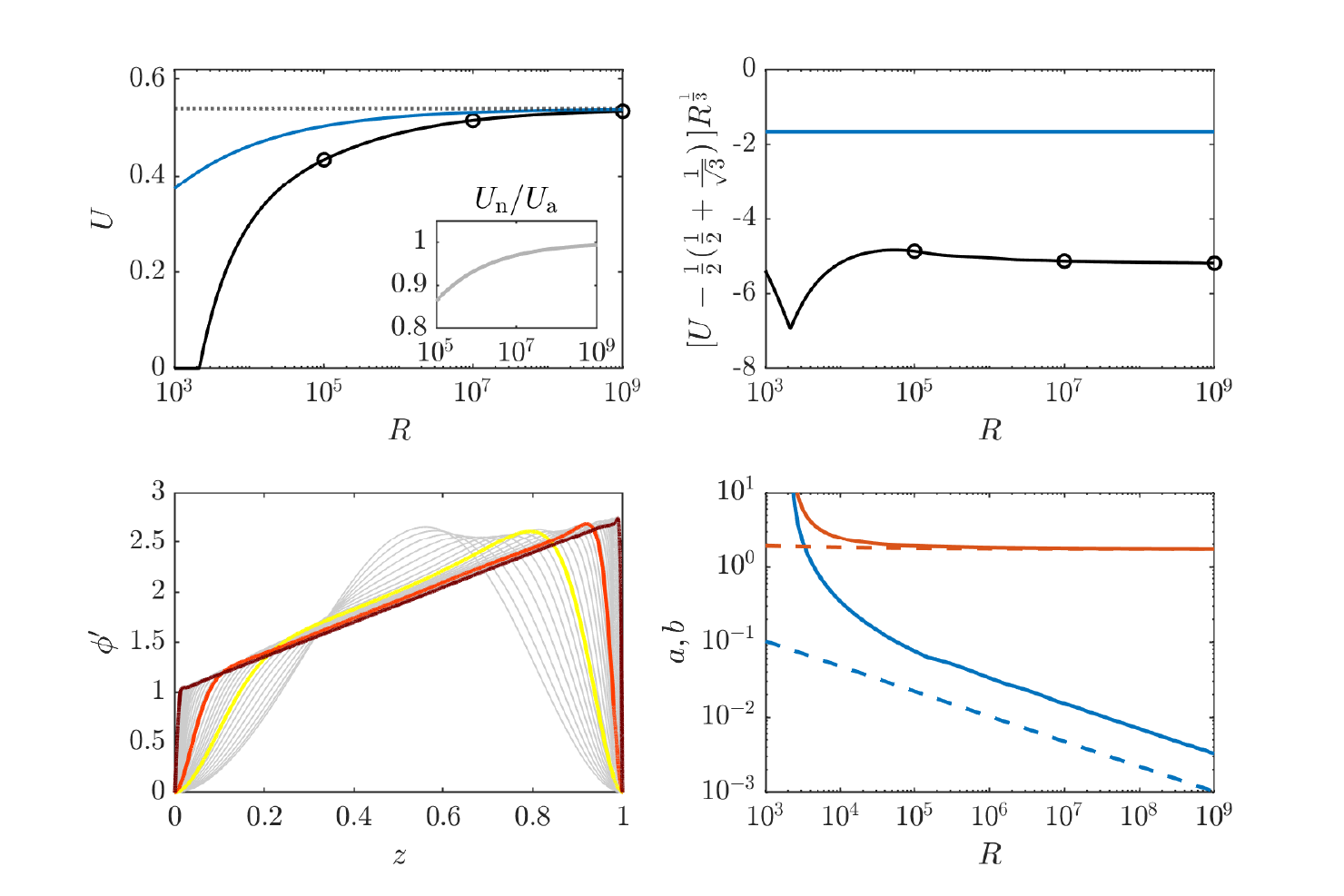}
    \label{fig:a_and_b_new_variations}
    \begin{tikzpicture}[overlay]
        \node at (-5.4,9.35) {\textit{(a)}};
        \node at (-1.2,6.9) {\textit{(b)}};
        \node at (5.7,9.35) {\textit{(c)}};
        \node at (-5.4,4.6) {\textit{(d)}};
        \node at (5.7,4.6) {\textit{(e)}};
        \node at (6.2,4.2) {\scriptsize $\sqrt{3}$};
        \draw[->,black!90] (-3.2,4.5) -- (-2.8,3.2) node[anchor=north] {\scriptsize Increasing \Ra};
    \end{tikzpicture}
    \vspace*{-2ex}
    \caption{
    \textit{Panel (a)}: Numerically optimal bounds $U_{\rm n}$ computed with \quinopt (\solidrule), compared to the analytical bound $U_{\rm a}$ \eqref{e:explicit_analytic_bound} ({\color{matlabblue}\solidrule}) and the improved uniform upper bound $\frac12(\frac12 + \frac{1}{\sqrt{3}})$ (\dottedrule). Insert \textit{(b)} shows the ratio of the two \Ra-dependent bounds.
     \textit{Panel (c)}: Analytical ({\color{matlabblue}\solidrule}) and numerical (\solidrule) corrections to the uniform bound $\frac12(\frac12 + \frac{1}{\sqrt{3}})$, compensated by $\Ra^{\frac13}$.
    \textit{Panel (d)}: Numerically optimal profiles $\phi'$ for $10^3 \leq \Ra \leq 10^9$ ({\color{mygrey}\solidrule}). Highlighted profiles for $\Ra = 10^{5}$ ({\color{colorbar1}\solidrule}), $\Ra = 10^{7}$ ({\color{colorbar10}\solidrule}) and $\Ra = 10^{9}$ ({\color{colorbar17}\solidrule}) correspond to the circles in panels \textit{(a)} and \textit{(c)}.
    \textit{Panel (e)}: Balance parameters $a$ ({\color{matlabblue}\solidrule} optimal; {\color{matlabblue}\dashedrule} analytical) and $b$ ({\color{matlabred}\solidrule} optimal; {\color{matlabred}\dashedrule} analytical).
    }
    \label{fig:psi_a_and_b}
\end{figure}

The numerically optimal bounds on $\langle w T \rangle$ are compared to the analytical bound \eqref{e:explicit_analytic_bound}  in Figure~\ref{fig:psi_a_and_b}\textit{(a)}. The former are zero until $\Ra_E = 2\,147$, which differs from the energy stability limit reported by \cite{Goluskin2015} due to the choice of horizontal period made in our numerical implementation. Insert $\textit{(b)}$ reveals that the optimal and analytical bounds appear to tend to the same asymptotic value as $\Ra \to \infty$. Moreover, as evidenced by panel $\textit{(c)}$, they seem to do so at the same rate. This suggests that the only possible improvement to our analytical bound is in the coefficient of the $O(\Ra^{-\frac13})$ correction, which for our numerically optimal bound is estimated to be $5.184 \pm 0.062$.

Figures~\ref{fig:psi_a_and_b}\textit{(d)} and \ref{fig:psi_a_and_b}\textit{(e)} show the optimal profiles of $\phi'$ and the optimal balance parameters $a,b$ in the range of $\Ra$ spanned by our computations. For large $\Ra$, the optimal $\phi'$ are approximately piecewise linear, corroborating our analytical choice in \eqref{eq:phi_g},  and the optimal balance parameters behave like the analytical ones from \S\ref{sec:explicit_ra_bound} (plotted with dashed lines). The main differences between the optimal and analytical $\phi'$ profiles are oscillations near the edge of the boundary layers and the fact that the two boundary layers of the optimal $\phi'$ have different widths. This suggests that a better prefactor for the $O(\Ra^{-\frac13})$ term in our analytical bound could be obtained, at the expense of more complicated algebra, by considering boundary layers of different size.

\section{Conclusions}\label{sec:conclusion}
\noindent
We have proved that the mean convective heat transport $\langle wT \rangle$ in IH convection with fixed boundary flux is rigorously bounded above by $\tfrac{1}{2}\big(\tfrac{1}{2}+ \tfrac{1}{\sqrt{3}} \big)  - cR^{-\frac13}$ uniformly in $\Pran$, where $c \approx 1.6552$. This result is the first to depend explicitly on the Rayleigh number and halves the previous uniform bound $\langle wT \rangle\leq \frac12 + \tfrac{1}{\sqrt{3}}$ \citep{goluskin2016internally} in the infinite-\Ra\ limit. Our proof relies on the construction of a feasible solution to a convex variational problem, derived by formulating the classical background method as the search for a quadratic auxiliary function in the form~\eqref{eq:aux_functional}. Numerical solution of this variational problem yields bounds that approach the same asymptotic value as \Ra\ increases and, crucially, appear to do so at the same $O(\Ra^{-1/3})$ rate. This suggests that our analytical bound is qualitatively optimal within our bounding approach, the only possible improvement being a relatively uninteresting increase in the magnitude of the $O(\Ra^{-1/3})$ correction to the asymptotic value. In particular, we conclude that the background method (at least, as formulated here) cannot prove that $\langle wT \rangle \leq \frac12 - O(\Ra^{-1/3})$ as conjectured by \citet[\S1.6.3.4]{goluskin2016internally}.

With the identity \eqref{e:J_flux}, our upper bound on $\langle wT \rangle$ can be translated into the lower bound $ \overline{T}_0 - \overline{T}_1  \geq \tfrac{1}{2}(\tfrac{1}{2}- \tfrac{1}{\sqrt{3}} ) + 1.6552\, R^{-\frac13} $. This bound is negative when $\Ra \geq 78\,390$, so conduction downwards from the top to the bottom cannot be ruled out in this regime. Determining whether $\overline{T}_0 - \overline{T}_1$ can indeed be negative or is positive at all Rayleigh numbers, as physical intuition suggests,  remains an open question for future work. Possible approaches to answer this question include direct numerical simulations at high $\Ra$, the construction of incompressible flows with optimal wall-to-wall transport \citep{Hassanzadeh2014,Tobasco2017,doering2019optimal}, and the computation of certain steady solutions to the Boussinesq equations~(\ref{e:governing-equations}\textit{a--c}), which in Rayleigh--B\'enard convection have been shown to transport heat more efficiently than turbulence over a wide range of $\Ra$ \citep{WenSteady2020}. These and other alternatives could provide a crucial understanding of the difference between the conjectured $\frac12$ asymptotic value for $\langle wT \rangle$ and the larger asymptotic value, $\frac12(\frac12 + \tfrac{1}{\sqrt{3}})$, of the upper bound proved in this paper.

\small
\vspace{2ex}\noindent
\textbf{Funding}  A.A. acknowledges funding by the EPSRC Centre for Doctoral Training in Fluid Dynamics across Scales (award number EP/L016230/1). G.F. was supported by an Imperial College Research Fellowship.

\vspace{2ex}\noindent
\textbf{Conflict of interests} The authors report no conflict of interests.

\vspace{2ex}\noindent
\textbf{Author ORCIDs}\\
Ali Arslan, {\color{matlabblue}https://orcid.org/0000-0002-5824-5604};\\
Giovanni Fantuzzi, {\color{matlabblue}https://orcid.org/0000-0002-0808-0944};\\
John Craske, {\color{matlabblue}https://orcid.org/0000-0002-8888-3180};\\
Andrew Wynn, {\color{matlabblue}https://orcid.org/0000-0003-2338-5903}.